\documentclass[10pt,prb,nobibnotes,showpacs,showkeys,superscriptaddress,maxnames=6]{revtex4-1}
\usepackage{geometry}
\usepackage{amsfonts,amsmath,amssymb,graphicx, placeins, bm}
\usepackage{setspace,color}

\usepackage{ae,aeguill}
\usepackage[utf8]{inputenc}    
\usepackage[T1]{fontenc}
\newcommand{\ddi}{\hat{\mu}_i\hat{\mu}_j -3 (\hat{\mu}_i \hat{r}_{ij})(\hat{\mu}_j \hat{r}_{ij})}

  \newcommand{\be}{\beta}     
\newcommand{\la}{\lambda}    
\newcommand{\gD}{\Delta}  \newcommand{\ep}{\epsilon} \newcommand{\s}{\sigma}
\newcommand{\om}{\omega}     
 \newcommand{\tend}{\rightarrow}
\newcommand{\equa}[1]{\begin{eqnarray} \label{#1}}
\newcommand{\auqe}{\end{eqnarray}}
\newcommand{\equab}[1]{\begin{widetext}\begin{eqnarray} \label{#1}}
\newcommand{\auqeb}{\end{eqnarray}\end{widetext}}
\newcommand{\tab}[1]{\begin{tabular}{#1}}
\newcommand{\bat}{\end{tabular} \\ }
\newcommand{\blanc}{\hskip 0.10\textwidth}
  
\begin {document}
  \title {Optimizing the magnetic hyperthermia efficiency from the structural dependence
          of dipolar interactions in magnetic nanoparticles assemblies.}
  \author {A. Morjane}
  \affiliation {PROMES-CNRS (UPR-8521), Université de Perpignan Via Domitia,
  Rambla de la thermodynamique, Tecnosud, 66100 Perpignan, France. }
  \affiliation{ICMPE, UMR 7182 CNRS and UPE 2-8 rue Henri Dunant 94320 Thiais, France.}
  \author {F. Vernay}
  \email [e-mail address: ] {francois.vernay@univ-perp.fr}
  \affiliation {PROMES-CNRS (UPR-8521), Université de Perpignan Via Domitia,
  Rambla de la thermodynamique, Tecnosud, 66100 Perpignan, France. }
  \author {V. Russier}
  \email [e-mail address: ] {vincent.russier@cnrs.fr}
  \affiliation{ICMPE, UMR 7182 CNRS and UPE 2-8 rue Henri Dunant 94320 Thiais, France.}
  \begin {abstract}
   \vskip 0.05\textheight
 \label {abstract}
 Magnetic hyperthermia, based on the conversion of the magnetic energy of magnetic
nanoparticles (MNP) under an alternating external field into heat,
presents a great interest, especially because of important applications
such as promising advances in cancer treatment.  
The underlying frequency-dependent response of the MNP
assembly under the excitation field depends on the first hand on the individual
properties of the MNP such as their size, shape and magnetocrystalline anisotropy
and on the other hand on collective properties generated by interactions between MNP. 
Due to the long-range dipolar component of the latter, the first consequence
is the so-called demagnetizing effects, making the response dependent on the external
sample shape. 
This is the reason for studying the properties of MNP in assemblies of anisotropic 
external shapes such as films, chains, or elongated aggregates. 
In the present work, using a recently developed Time Quantified Monte Carlo (TQMC) framework for
dynamic simulations, we focus more precisely on the importance of the local
structure in such anisotropic aggregates to the dipolar interactions' contribution 
of the response to the external field.
Here the local structure is understood as both the location of the nanoparticles and
the distribution of their easy axes. We show that the latter must be determined in a self-consistent 
way starting with the equilibrium structure of a ferrofluid modeled as
the dipolar hard sphere fluid.
  \end {abstract}
  \maketitle
%
 \section {Introduction.}
 \label {intro}
 Dynamic properties of magnetic nanoparticle (MNP) assemblies play an important role in the field 
of nanoparticle research and especially for hyperthermia applications, which consist of the conversion into 
heat of the magnetic energy absorbed by MNP under an external alternating magnetic field (AMF).
 The latter generates a great deal of 
interest, essentially due to applications in nanoscale medicine, especially for 
treatments against cancer~\cite{pankhurst_2009,gavilan_2021b,baldea_2025} or 
catalysis~\cite{bordet_2016,bordet_2025}. In the former case, magnetic hyperthermia is expected 
to provide very localized heating in the vicinity of the tumor where MNP are injected to 
either directly damage cancer cells or increase the localized therapeutic action through the temperature increase.

However, dispersed MNP cannot provide a significant temperature 
increase, and one must deal with dense MNP assemblies or 
aggregates~\cite{hergt_2006a,keblinski_2006,eggeman_2007}.
Moreover, the heat dissipation processes at the MNP interface, across the embedding medium 
and at the sample-environment interface must be taken into account for the whole heating 
process~\cite{dejardin_2022,kachkachi_2026}. 
Hence, beyond very small concentrations of MNP, the question of the 
influence of the MNP interactions on the efficiency of the hyperthermia through the 
specific absorption rate (SAR) is of crucial importance, at least to provide a MNP interaction
dependent source term beyond the weak coupling regime where the perturbation 
description holds~\cite{jonsson_2001,azeggagh_2007}.
In the case of immobilized MNP, the dominant interaction between neighboring MNP is the dipole-dipole 
interaction (DDI) since one refers to MNP in the single magnetic domain regime.
It is thus of crucial importance to determine precisely the DDI effect on the SAR
not only with respect to the MNP concentration in the assemblies but also in terms of both
their external shape and local structure.

The effect of inter-particle interactions on the SAR in concentrated assemblies of MNP
has been addressed already for a long time~\cite{haase_2012,serantes_2010} 
together with the influence of the external shape
in relation to the demagnetizing effects.
In particular, it is now well recognized that in isotropic assemblies the DDI decreases the 
SAR~\cite{serantes_2010,haase_2012}.
In order to conversely improve the SAR with respect to the DDI, it is necessary to consider
anisotropic-shaped aggregates such as film-like or prolate ones~\cite{mehdaoui_2013,gudoschnikov_2012a,gudoschnikov_2012b}
due to demagnetizing effects in a first approximation. 
On the other hand, it has been evidenced that MNP assemblies can self-organize
in elongated or needle-shaped aggregates under the action of a field even in
hyperthermia conditions~\cite{mehdaoui_2013,sanz_2020,morales_2021,mille_2021}. 
In these experimental works, a strong enhancement of the hysteresis loops 
results from the chain formation.
Similarly, elongated MNP aggregates have been elaborated by template-like methods~\cite{gudoschnikov_2012a,arica_2026}
in order to increase the SAR value and the natural chain arrangements of cubic MNP
or magnetosomes in magnetotactic bacteria are also found to increase the SAR compared to the same 
magnetosomes in isotropic assemblies~\cite{gandia_2019}.
This strongly motivates the study of the effects of the DDI
on the SAR or dynamic properties of chain-like and elongated-shaped MNP aggregates.
In such geometries, when the AMF is along the major axis, the general finding is an increase of the
system's magnetic response when compared to the isolated MNP. 
However these works corresponds to either 
linear chains~\cite{martinez-boubeta_2013,anand_2019,anand_2020,anand_2022}
or MNP located on a regular 
lattice~\cite{mehdaoui_2013,saville_2014,sanz_2020}.
In Ref.~\onlinecite{schoenen_2025}, the MNP positions in the elongated aggregates are obtained by an
advancing front method leading to a disordered but isotropic hard-sphere-like structure. 

In this work, we perform dynamic Monte Carlo (MC) simulations~\cite{morjane_2025} to study
the effect of the DDI on the SAR of aggregates of MNP, focusing especially
on their shape and local structure. 
We are interested in a system of immobilized MNP free of super-exchange interactions 
that we model by an ensemble of dipolar hard spheres with uniaxial anisotropy.
The local structure in the aggregates is expected to be of crucial importance for
the response to the AMF and thus to the SAR 
through the DDI-induced anisotropy. Indeed, the latter has been found to favor the 
ferromagnetic behavior in the bulk dipolar hard sphere fluid (DHS)~\cite{malherbe_2023}.
Here, the structure is understood not only as the location of the MNP
inside the clusters but also as the distribution of the MNP easy axes, which is found to play an important role.
More precisely, we show that the self-consistent determination of the latter, given the equilibrium
structure determined from the DHS in the aggregate, must be considered.

The paper is organized as follows. In section~\ref{model} we present the model, the simulation
methods and the determination of the local structure are presented in section~\ref{method}. 
Then the results are given in section~\ref{results} before concluding remarks in 
section~\ref{conclusion}.
 \section {Model}
 \label   {model}
 The model is an assembly of $N$ spherical MNP in the single-domain regime, free of super-exchange interactions and 
characterized by a uniaxial magnetocrystalline anisotropy energy (MAE) and monodisperse in size with diameter $d$ and 
volume $v$.
 We place ourselves in the framework of the effective one-spin model where each single-domain MNP is assumed to be
uniformly magnetized with a temperature-independent saturation magnetization $M_s$.
 Hence, we
consider a system of dipolar hard spheres of moment $\vec{\mu}_i=\mu\hat{\mu}_i$, with $\mu=M_{s}v$, 
interacting through the usual dipole-dipole interaction (DDI), subjected to a 
one-body anisotropy energy, $K_{u}v(\hat{n}_i\ldotp\hat{\mu}_i)^2$ and to an external field 
$\vec{H}=H_0\hat{h}$ where $H_0$ may be time dependent.
$K_u$ and $\hat{n}_i$ are the anisotropy constant and the easy axis of the particle $i$, respectively.
We introduce a reference temperature, $T_0$ ($\be_0=1/k_BT_0$) and the reduced temperature $T^*=T/T_0$.
The Hamiltonian of the system is then given by
 \equa{beh_1}
  \be H  &=&  \frac{1}{T^*} \left( \frac{\la_d}{2} \sum_{i\neq j} \frac{\ddi}{(r_{ij}/d)^3} 
            - \la_u \sum_i (\hat{n}_i\ldotp\hat{\mu}_i)^2  - \frac{1}{2}\be_0 \sum_{i\neq j} v_{sr}(r_{ij})
            - h \sum_i \hat{h}\ldotp \hat{\mu}_i  \right )
  \\ \nonumber
    & \la_u & = \be_0 K_uv(d) ~;~ \la_d = \be_0\frac{\mu_0}{4\pi}\frac{\mu^2}{d^3} ~;~ h = \be_0\mu_{0} \mu H_{0} ~\equiv~ 2\la_{u}\frac{H_0}{H_K}  
 \auqe
where $\hat{r}_{ij}$ is the unit vector carried by the vector joining sites $i$ and $j$,
$\be=1/(k_BT)$ is the inverse temperature.
$v_{sr}(r)$ is the short-range contribution, taken as the hard-sphere potential, 
$v_{sr}(r>d)=0$ and $v_{sr}(r<d)=\infty$.
Equation~(\ref{beh_1}) introduces the 
DDI, MAE and Zeeman coupling constants, $\la_d$, $\la_u$ and $h$,
where $H_K=2K/\mu_{0}M_s$ is the usual anisotropy field.
The distribution of easy axes, $\{\hat{n}_i\}$ is either fixed or self-consistently deduced from the equilibrium
configuration of the system (see section~(\ref{struct})). In the former case, we choose a Gaussian-like distribution
of the polar angle $\theta$ of the $\hat{n}$,
\equa{dist_n}
 P(\theta) = C\frac{\sin(\theta)}{2}\left[\exp(-\frac{\theta^2}{2\s^2}) + \exp(-\frac{(\theta-\pi)^2}{2\s^2})\right],
\auqe
where the random distribution, $P(\theta)=\sin(\theta)/2$ is recovered in the limit $\s\gg\pi$.
\\
In this work, we consider mainly finite-size aggregates with free boundary conditions, except in the film geometry,  
where periodic boundary conditions are used in the 2 directions parallel to the film and then the 
Ewald summation technique is used~\cite{allen_1987,weis_1993} for the long-range DDI interaction.
%
\section {Method}
\label {method}
\subsection {Monte Carlo simulations and observables}
\label {mc}
 Two kinds of MC simulations are considered, concerning either the determination of equilibrium 
structures, for which we use a parallel tempering scheme~\cite{earl_2005}
or the simulation of dynamical susceptibilities and hysteresis, performed
according to the method we developed~\cite{morjane_2025}
from the so-called TQMC scheme~\cite{nowak_2000,cheng_2006}.
For the latter, we refer the reader to these works and do not enter into the details. 
We just need to recall that it is based upon the precise tuning of the aperture of the cone, $R$,
in which the new moment at site $i$ say $\hat{\mu}_i^{(n)}$ is chosen from 
its preceding direction, $\hat{\mu}_i^{(o)}$. 
Rewriting $R=R_0\sqrt{T^*}$ the unit of time corresponding to $\gD{}N$ MC steps can be introduced
 \equa{dt3}
   u_t = \bar{\Gamma}_{acc}\gD{}N R_0^2
 \auqe
where $\bar{\Gamma}_{acc}$ is the average value of the acceptance rate of the MC paths in the range of temperatures considered.  
An important point is that equation~(\ref{dt3}) is valid for both the Metropolis and the heat-bath sampling.
In the following, the dynamical MC simulations concern only ensembles of fixed MNP, where the moments are the 
only degrees of freedom.
\\ \indent
From this point, we have to distinguish between the instantaneous MC time step $i$, which is nothing but the index of 
MC step (as usual one MC step corresponds to a trial move for all the moments) and the MC time $t_{MC}$, 
which includes the unit of time as defined by equation~(\ref{dt3}), namely $t_{MC}=u_{t}(i/\gD{}N)$. 
%
In most of our simulations, especially for the relaxation process, we divide the total number of MC steps, say $N_m$,
into intervals of $\gD{}N$ playing the role of a finite $\gD{}t$ time interval. Then we consider sets of $\gD{}N$ steps 
which will be labeled by $i_{MC}\;\in[1, N_m/\gD{}N]$ and define the MC time
as $t_{MC}=i_{MC}u_t$. \\ \indent
The observables considered in this work are, besides the total energy, on the one hand, the magnetization $M(t_{MC})$ 
projected on the direction of the external field $\hat{h}$
as the thermal average of the instantaneous $m(i)$ over $\gD{}N$ MC steps and
a large number (100 or 200) of realizations of the system, as the studied
properties are ''time dependent'':
\equa{M_t1}
 M(t_{MC}) = \langle m(i_{MC})\rangle_{\gD{}N} = \frac{1}{\gD{}N}\sum_{i=i_{MC}\gD{}N}^{(i_{MC}+1)\gD{}N} m(i)
  \blanc \textrm{with} ~~ m(i) = \frac{1}{N}\sum_{k=1}^N \hat{\mu}_k\ldotp\hat{h}.
\auqe
On the other hand, the susceptibility $\chi(\om)$ in terms of the frequency $\om$ is 
determined from the method introduced by Andersson {\it et al.}~\cite{andersson_1997}. 
For the latter, the external reduced field is taken as $h(t)=h_0sin(\om{t})$ and we have
\equa{chi1}
        \chi'  = \frac{2}{h_0 N_{p}} \sum_{i=1}^{N_p} \sin(\frac{2\pi}{N_p}i) m(i)
 \blanc \chi'' = \frac{2}{h_0 N_{p}} \sum_{i=1}^{N_p} \cos(\frac{2\pi}{N_p}i) m(i)
\auqe
where $N_p$ is the number of MC steps corresponding to one period, {\it i.e.}
$N_p.u_t=2\pi/\om$. The period corresponds to the measuring time $t_m=N_p u_t$. From a practical 
point of view, the whole MC run involves $N_m$~MC steps covering a number $p$ of periods; 
the averages involved in equation~(\ref{chi1}) are performed over the last $p/2$ periods.
and the corresponding hysteresis loops are recorded only on the last one.
Equation~(\ref{chi1}) is written for $\gD{}N=1$ for the sake of simplicity, but can easily be translated to the more 
general case of $\gD{}N>1$ as 
\equa{chi2}
      \chi'  = \frac{2}{h_0 N'_{p}} \sum_{i=1}^{N'_p} \sin(\frac{2\pi}{N'_p}i)\langle m(i\gD{}N)\rangle_{\gD{}N}
      ~ \textrm{with} ~ N'_p = \frac{N_m}{\gD{}N}\frac{1}{p}
\auqe
and similarly for $\chi''$. In any case, when translated to the {\it actual} MC time, the value of $\gD{}N$ does not play any role. 
It is worth mentioning that the susceptibility as written in equation~(\ref{chi1}) or~(\ref{chi2}), is the first harmonic susceptibility
denoted as $\chi^{''}_1$ in the following,
corresponding to the coefficients $a^{'}_1(h_0)$ and $a^{''}_1(h_0)$ of the Fourier series of the magnetization~\cite{barrera_2023},
\equa{m_t}
   m(t) = \sum_{n=0}^{\infty} \left\{ a^{'}_{2n+1} \sin\left[(2n+1)\om t\right] + a^{''}_{2n+1} \cos\left[(2n+1)\om t\right] \right\}.
\auqe
The efficiency of the hyperthermia process is measured by the specific absorption rate
(SAR)~\cite{rosensweig_2002} related to the area of the hysteresis loop over one period of the AMF, 
given in reduced units and relative to one MNP by %
 \equa{sar}
   SAR = \frac{\om{}h_0^2}{2} \chi^{''}_1 = \frac{\om{}h_0}{2} a^{''}_1
 \auqe
Here $\chi^{''}_1$ corresponds to our reduced units and is related to the actual dimensionless one $\tilde{\chi}^{''}_1$ by 
\equa{chi2_dim}
 \tilde{\chi}^{''}_1 = \frac{H_0}{M_s}\chi^{''}_1
\auqe
The frequency $\om$ corresponds to $2\pi/u_t$ and is therefore unitless. To make the link with actual dimensioned quantities, we
first recall that we can relate $u_t$ to the actual time that it represents through the temperature at the maximum of 
$\chi^{'}(\om)$, $T_m(\chi^{'})$, as is done in Ref.~\onlinecite{morjane_2025} once the value of $T_0$ is chosen. 
The method consists in adjusting the number of MC steps $\gD{}N$ to represent $\om$ by equating the experimental value of the 
temperature $T_m(\chi^{'})$ to the simulated one, $T_0T_m^*$. 
Examples were given in Ref.~\onlinecite{morjane_2025} for maghemite MNP of diameter 8~$nm$ (see Ref.~\onlinecite{de-toro_2013b}) 
and 15 $nm$ (see Ref.~\onlinecite{garcia-acevedo_2023}) with $T_0=97$ and 318\;K respectively.
Then the SAR expressed in W/g is given by
\equa{sar_dim}
 SAR = \Phi \frac{k_{B}T_0}{2\pi v(d)\rho}\om h_0^2 \chi_{1\om}^{''}  
\auqe
for the system characterized by a magnetic volume fraction $\Phi$, 
where $\rho$ is the volumic mass of the MNP expressed in $g/m^3$ ($4.86\;10^6\;g/m^3$ for maghemite) and $v(d)$ 
is the MNP volume. In the following, we consider a set of aggregates differing both in shape and internal 
structure. A common feature of these is that they are built from disordered distributions of MNP with the 
same volume fraction, $\phi=0.45$, chosen to be reachable in MNP superstructures and far enough from the 
upper bound for hard spheres, the random close packing value, $\Phi_{RCP}=0.64$ 
which can be obtained only with uncoated MNP and after a sintering process~\cite{de-toro_2013b}.
%
%
\subsection {Structures of the MNP aggregates}
\label {struct}
  The shape and internal structures of the different aggregates used in this work
are motivated, on the one hand, to represent actual MNP aggregates and, on the other hand, to 
optimize the efficiency of the hyperthermia process through the effect of the DDI.

 These aggregates are characterized by both their external shape 
and internal structure. The external shape is either spherical, of film like geometry
or cylindrical with large length to diameter ratio which generalize the singular MNP chains of only one
MNP diameter width.
The local structure of the aggregates is either the hard sphere-like distribution, namely both disordered 
and isotropic or the dipolar hard sphere (DHS) fluid distribution, which may be anisotropic and present 
a preferential or nematic direction according to both the temperature and the external 
field~\cite{malherbe_2023}.

\begin {figure}
     \begin {center}
     \hskip -0.10\textwidth
     \includegraphics [width = 0.62\textwidth]{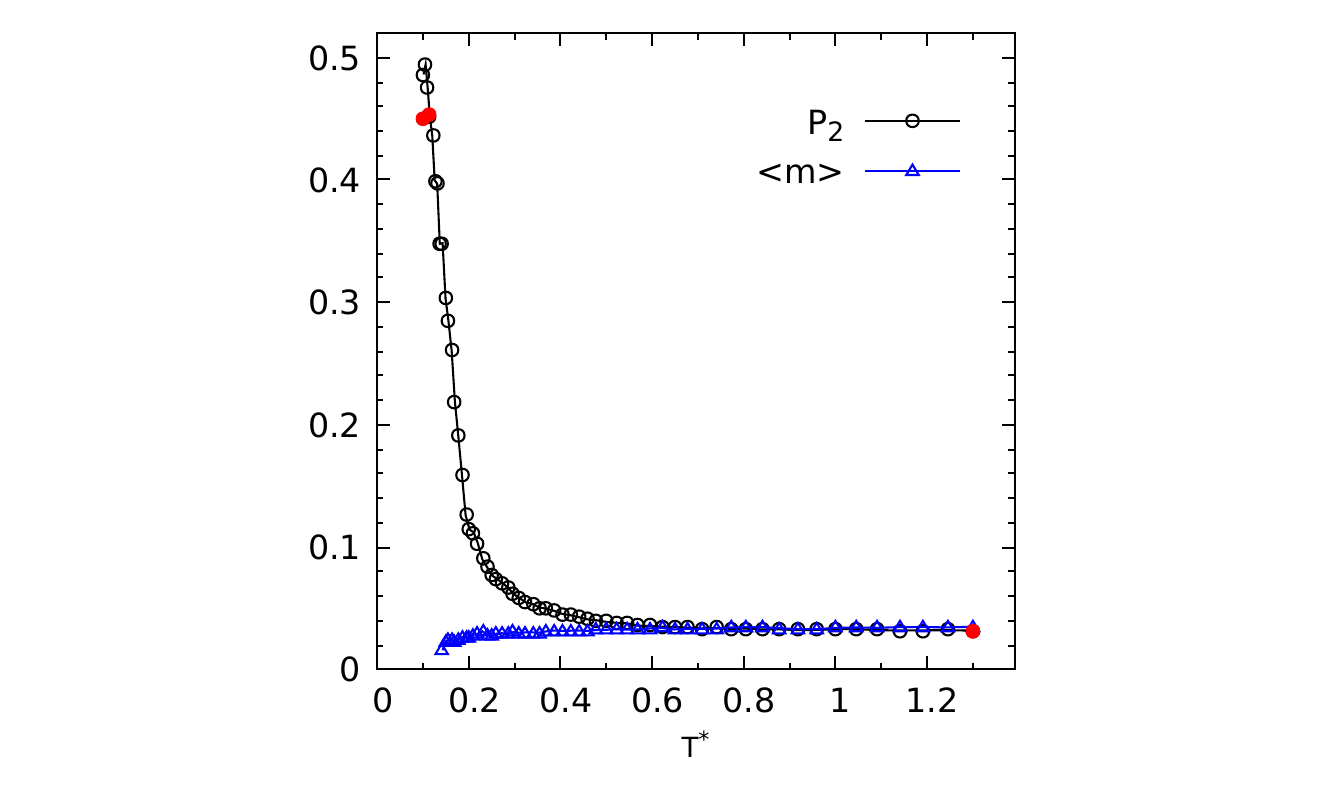}
     \hskip -0.15\textwidth
     \includegraphics [width = 0.62\textwidth]{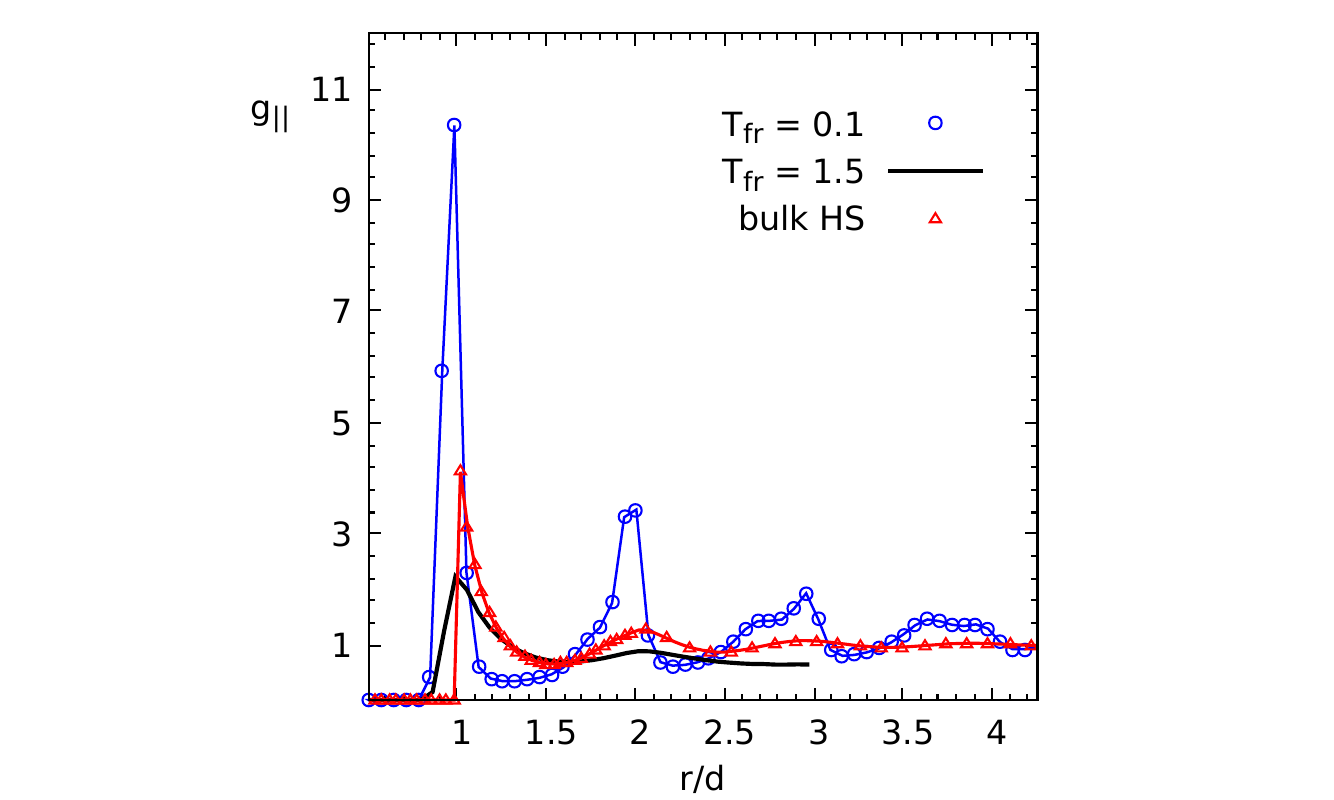}
     \end {center}
   \caption {\label {cyl_gpar}
   Left : Nematic order parameter $P_2$ and mean value of the magnetization, $<m>$
   of the DHS confined to the cylinder in the absence of an external field after 
   equilibrium obtained from parallel tempering MC. The red points at $T^*=0.1$
   and 1.30 correspond to the subsequent equilibrium states with 80 independent
   realizations starting from the initial parallel tempering configurations 
   of the full range of temperatures.
   Right: Corresponding pair distribution function in the direction parallel to the cylinder
   axis, $g_{\parallel}(r)$ at $T^*=0.1$ and 1.5 as indicated, compared the the bulk
   hard sphere pair distribution function at $\Phi=0.45$.
   }
  \end {figure}

\begin {figure}
   \includegraphics [width = 0.66\textwidth]{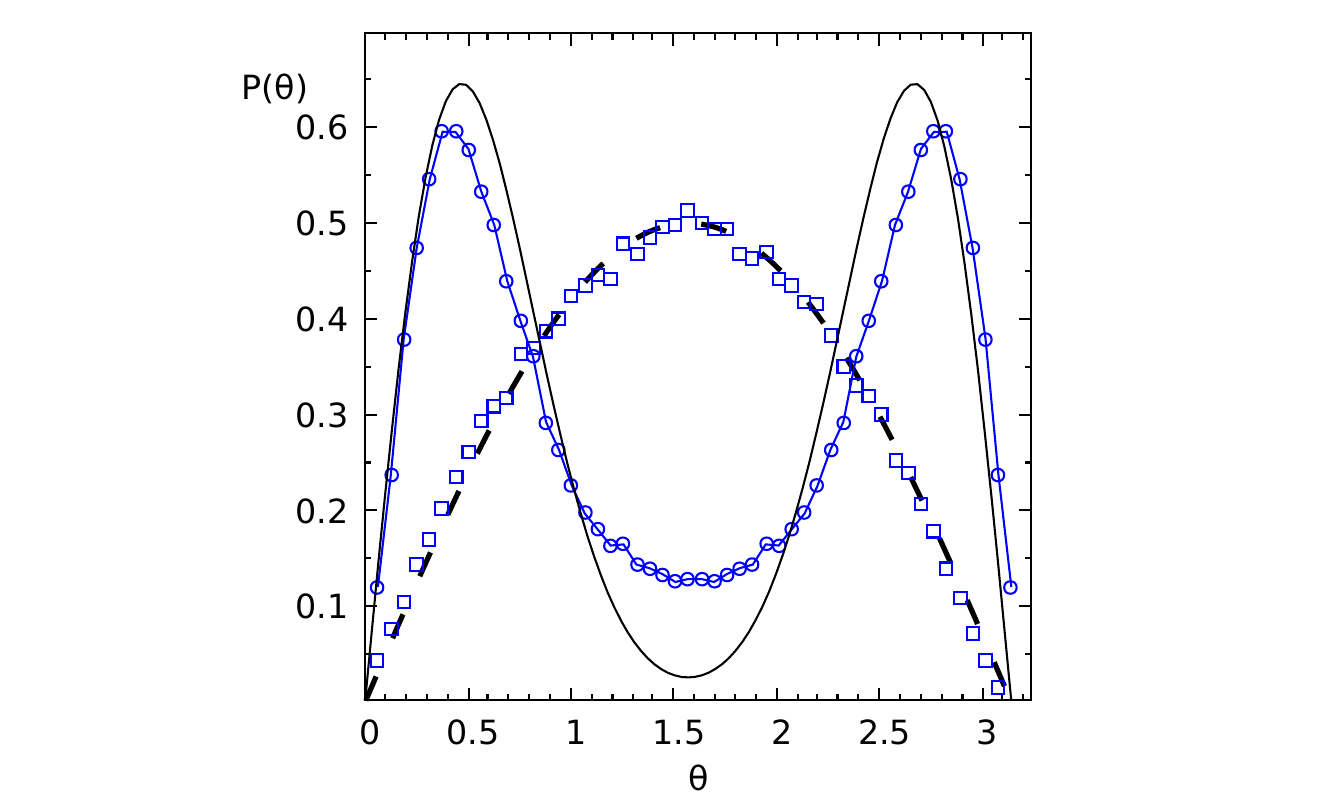}
   \caption {\label {histo_g4}
    Easy axes distribution probability of the DHS confined to the cylinder in the absence of an external field,
    relative to its principal axis at $T^*=0.11$, open circles and 1.50, open squares. 
    Dasched line : $P(\theta)=sin(\theta)/2$ correspnding to the random distribution; 
    solid line fit of $P(\theta)$ from the analytic form of equation~(\ref{dist_n}) with 
    $\s=0.489$, leading to $P_2=0.5$ from equation~(\ref{tens_q}).
   }
  \end {figure}

 The hard sphere like distribution is simply obtained by performing MC runs in the absence of DDI
and including $v_{sr}(r)=v_{hs}(r)$ in equation~(\ref{beh_1}) in a large cubic volume from 
which we cut the aggregate of chosen external shape. Here we choose either a sphere including c.a. 1000
MNP, a slab including c.a. 600 MNP, or an elongated cylinder. 
The ferrofluid distribution is obtained in two steps: we first consider the dipolar hard sphere fluid, with
only the hard-sphere and the DDI contributions in equation~(\ref{beh_1}), in the absence of MAE and
after equilibrium is reached, we impose the coincidence of the easy axes with the moments by setting 
$\hat{n}_i=\hat{\mu}_i$ on each site. 
This first step corresponds to the limit $\la_u\tend\infty$ and can be seen as the case of the Brownian rotation
in a weakly viscous solvent where the MNP rotate instantaneously as a whole.
We know that the DHS, even in the bulk, spontaneously polarizes and thus presents at low temperature a spontaneous 
nematic direction~\cite{levesque_2014,malherbe_2023}. The degree of anisotropy in the DHS is quantified by the nematic order
parameter, $P_2$, relative to the moments distribution, which is the largest eigenvalue of the tensor
 \equa{tens_q}
    \bar{Q} = \frac{1}{2N} \sum_i (3\hat{\mu}_i\hat{\mu}_i - \bar{I} ).
 \auqe
A non-vanishing value of $P_2$ is related to an anisotropy of the pair distribution function $g(r)$ displaying
a strong structuration in the nematic direction (see figure~(\ref{cyl_gpar})). 
Moreover, as has been shown in Ref.~\onlinecite{malherbe_2023} this structural anisotropy is also quantified by an
equivalent of the nematic order parameter $P_2$ relative to the nearest neighbors distribution, $\{\hat{l}_{nn}\}$
when the latter are inserted in equation~(\ref{tens_q})
instead of the $\{\hat{\mu}\}$ (see figure~(3) of~Ref.~\onlinecite{malherbe_2023}). This structural nematic order parameter is 
not shown in the present work.
Here, we specifically focus on the case of the DHS confined to a simulation box of cylindrical shape whose length and
radius are respectively $(L,R) = 67, 6.9$ in units of MNP radius. The spontaneous ordering takes place with the cylindrical
axis as the nematic direction, as expected from magnetostatics, and a vanishing total magnetization with the decrease 
of temperature as seen in figure~(\ref{cyl_gpar}). The angular distribution of the moments relative to the cylinder axis
as the polar axis is displayed in figure~(\ref{histo_g4}).

\begin {figure}
      \includegraphics [width = 0.65\textwidth]{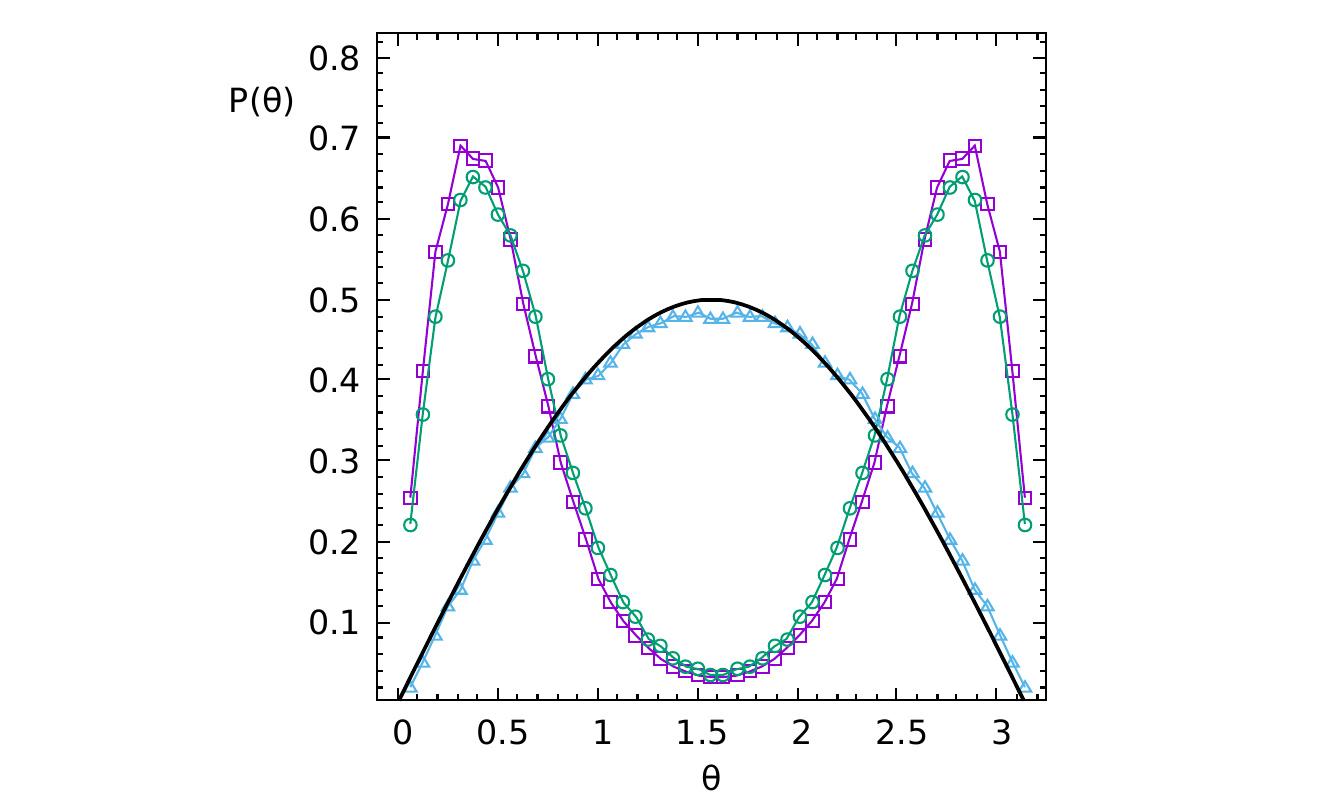}
    \caption {\label {dist_n_be316} 
              Easy axes distribution probability for the DHS with $h_{DC}=1$ frozen at $T^*_{fr}=0.316$
              and 2.0. Equilibrium~$\{\hat{n}\}$ at $T^*=0.316$, open circles;
              $\{\hat{n}\}_{\infty}$ at $T^*=0.316$, open squares.
              $\{\hat{n}\}$ at $T^*=2.0$, open triangles compared to the random distribution, solid line.  
              }
   \end {figure}

\begin {figure}
      \hskip -0.12\textwidth
      \includegraphics [width = 0.60\textwidth]{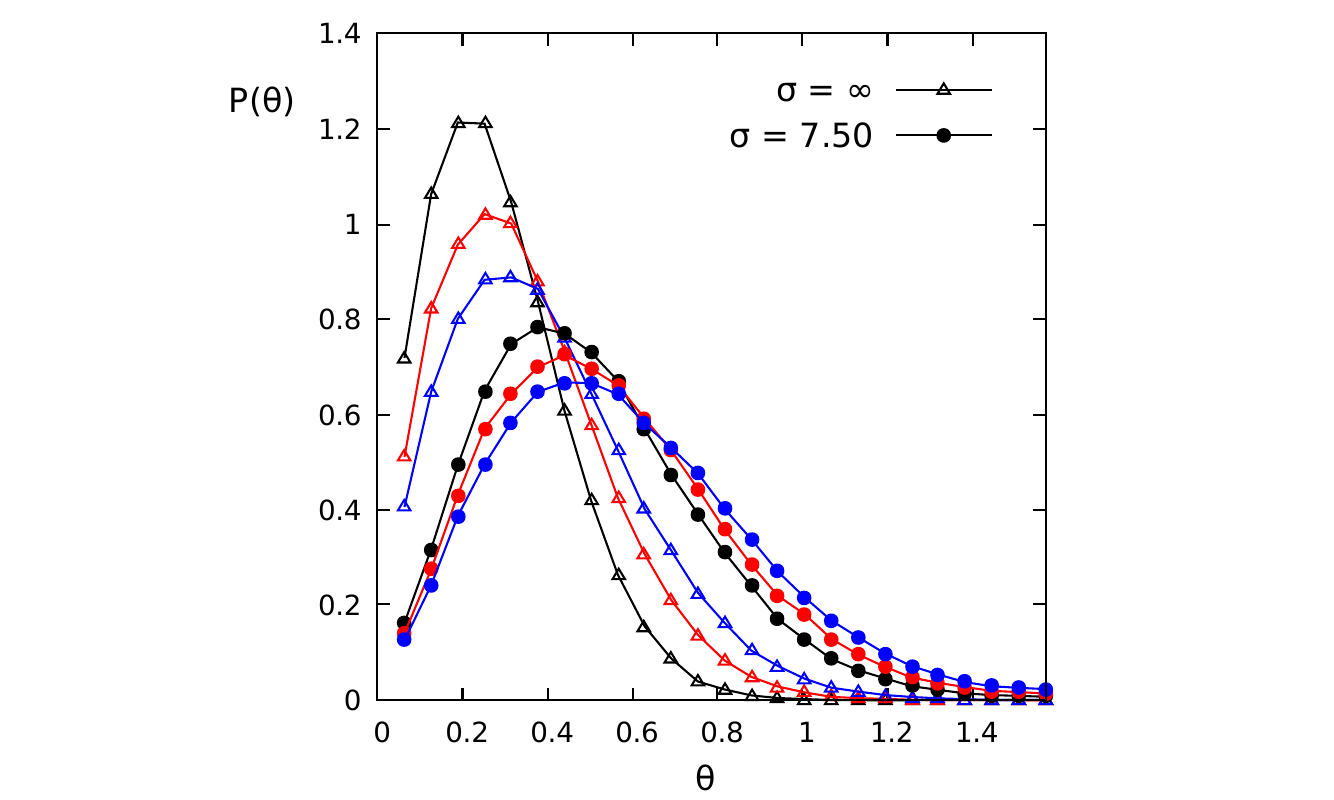}
      \hskip -0.12\textwidth
      \includegraphics [width = 0.60\textwidth]{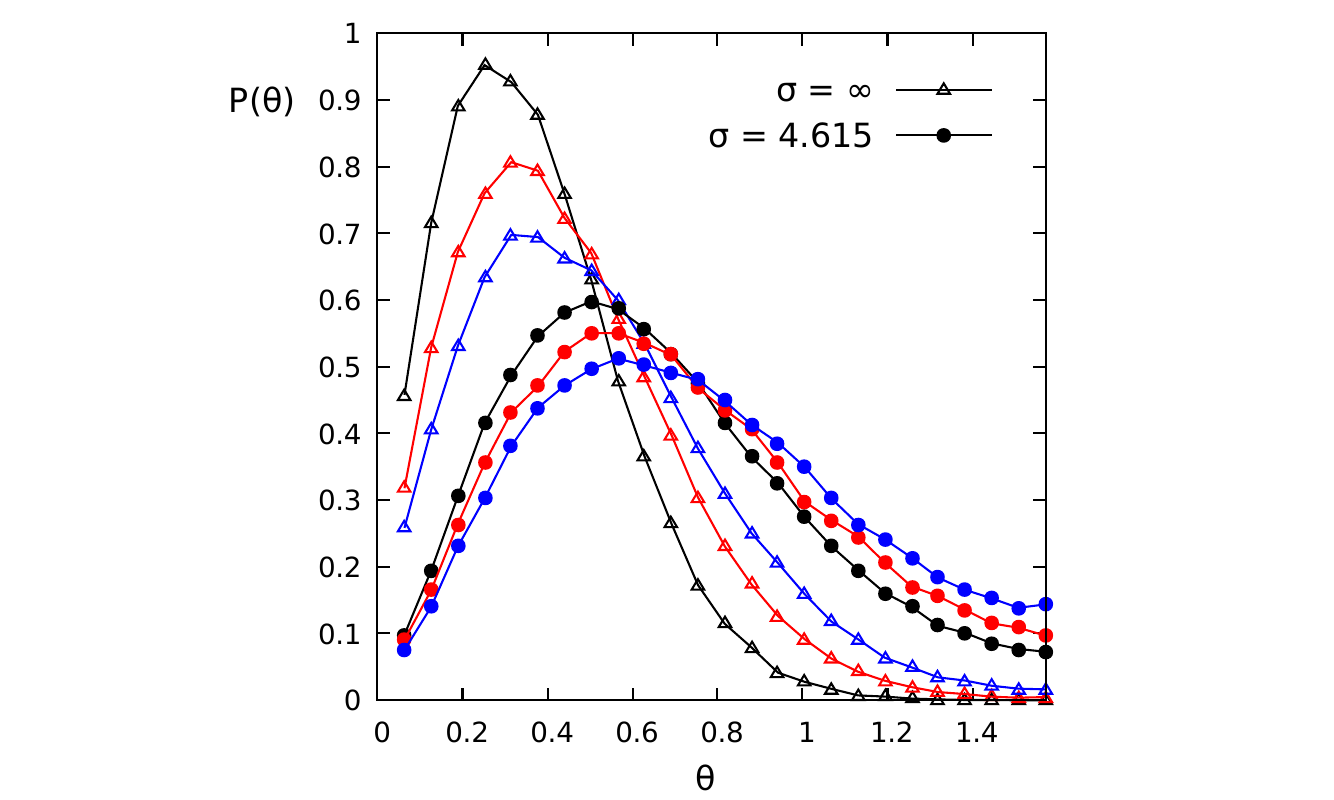}
    \caption {\label {dist_n_08-130}
      Same as figure~(\ref{dist_n_be316}) for the DHS at different values of the external field, 
     $h=12$, 8 and 6 from top to bottom.
      $\la_u=6$  and $T^*=0.8$, left, 1.30, right.
      $\s=\la_u/T^*$; $\s=\infty$ corresponds to $\{\hat{n}\}_{\infty}$.
    }
   \end {figure}

The second step consists in determining the easy axes distribution by taking into account the thermal 
activation~\cite{chantrell_2025}, namely to release the limit $\la_u\tend\infty$. To perform this step, 
we exploit the symmetry $(\hat{\mu},\hat{n})$ in the hamiltonian~(\ref{beh_1}) and determine 
the distribution $\{\hat{n}_i\}$ with the moments and MNP locations fixed at the DHS step configuration by 
a MC simulation using the finite value of $\la_u$ and $\la_d=0$ in equation~(\ref{beh_1}). 
Then the easy axes distribution depends on both $\la_u$ and temperature $T^*$.
We note that the correction brought in by the distribution of the easy axes is essentially quantitative
and becomes quite small at low temperature when $\s_u=\la_u/T*$ increases beyond $c.a.$~10. This is shown in 
figures~(\ref{dist_n_be316},\ref{dist_n_08-130}) where we compare the equilibrium $\{\hat{n}\}$ and the corresponding
$\{\hat{n}\}(\la_{u}\tend\infty)=\{\hat{n}\}_{\infty}$ 
for the DHS in presence of a DC field $h=1$ frozen at $T^*_{fr}=0.316$ (figure~(\ref{dist_n_be316})) 
and for the DHS in presence of a DC field $h=6$ to 12 and $T^*=0.8$ and 1.30 (figure~(\ref{dist_n_08-130})).

\begin {figure}
   \includegraphics [width = 0.66\textwidth]{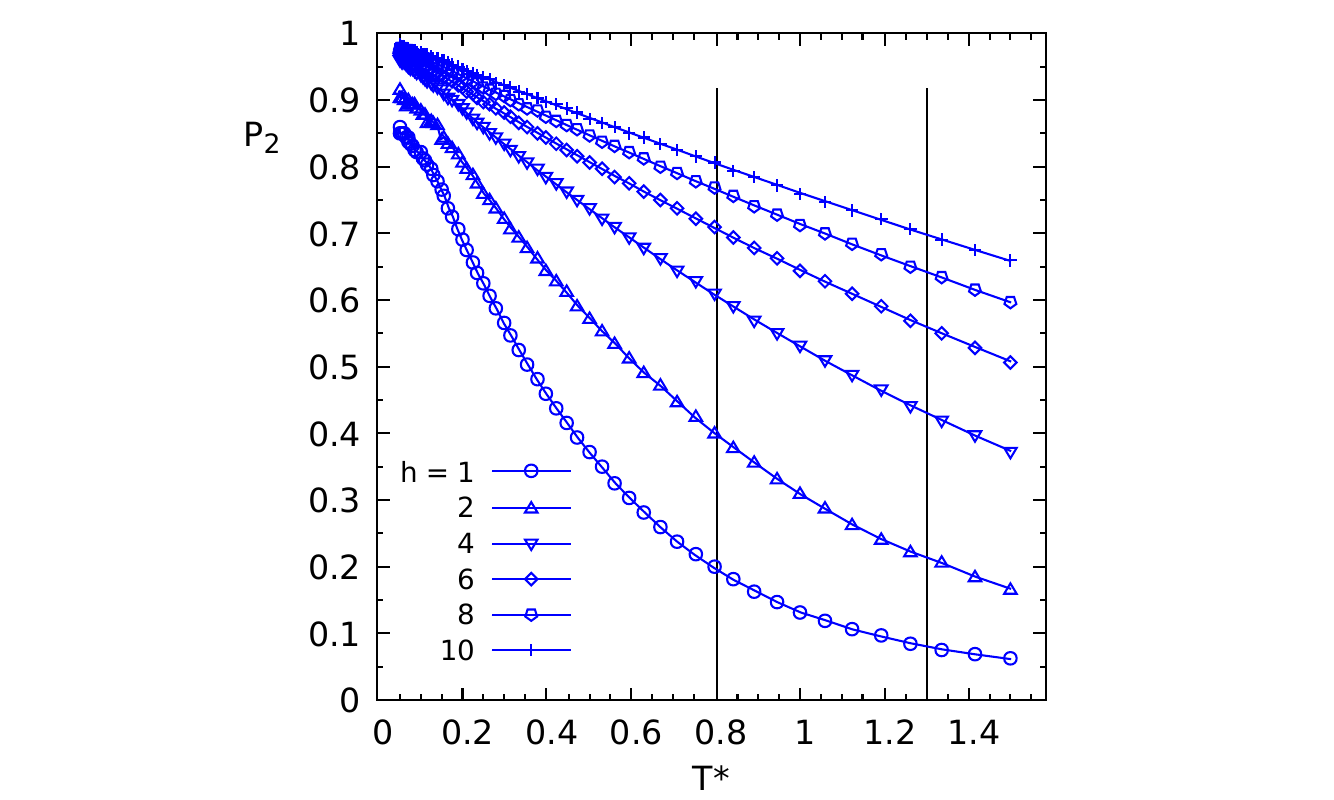}
   \caption {\label {lambda_h}
   Nematic order parameter $P_2$ of the cylindrical aggregate obtained from the DHS structure
   in the presence of an external DC field of magnitude $h$. The lines at $T^{*}=0.8$ and 1.3 are
   useful marks for the results of section~\ref{results}.
   }
   \end {figure}

The determination of the DHS structure, including the ($\la_u,T^*$) dependent distribution of the easy axes, can be
similarly performed in the presence of an external field $h_{DC}$. We consider here only the case of an external DC field
in the direction of the cylinder; this plays the role of the external field used to induce chain-like aggregates. 
In the presence of such a field, the latter increases the effect of the low temperature in the formation of the 
structure anisotropy, which means that a given degree of anisotropy as quantified by $P_2$ is obtained at a larger value 
of $T_f^*$ with $h_{DC}\neq{0}$ than with $h_{DC}=0$. 
The parameter $P_2$ is seen as an order parameter to quantify the degree of structural anisotropy for both cases
in the absence or in the presence of an external field. 
%
In the following, we will simulate the elongated aggregates obtained in 
either a DC or an AC field through the preceding steps by using a DC field along the direction of the cylinder.
Although the use of a DC field to simulate the structure of the aggregates formed in an AC field is approximate,
we think that the resulting structure is at least qualitatively correct.
The value of $P_2(h)$, displayed on figure~(\ref{lambda_h}), and accordingly the structure anisotropy 
depends strongly on the value of $h$.

Once the structure concerning both the location of the MNP $\{\vec{r}\}_i$ and their easy axes distribution $\{\hat{n}\}_i$
is obtained according to the above description, it is frozen, and we then perform the dynamic simulation in the framework 
of our TQMC scheme, in the AC field where the MNP moments $\{\hat{\mu}\}_i$ are the only degrees of freedom in order 
to study the first harmonic susceptibility and the hysteresis curves.
To get satisfactory statistics, for each structure we built $N_r$ different realizations with $N_r=100$
or 200. 

The susceptibility $\chi(\om)$ of the columnar aggregates depends first on their shape and second on the internal structure, 
both through the structural anisotropy, namely the difference between $g_{\perp}$ and $g_{\parallel}$ 
(see figure~(\ref{cyl_gpar}), and the texturation of the easy axes distribution. 
In order to quantify these two effects separately, we can fit the self-consistent distribution
 $\{\hat{n}\}_i$ on the analytical form, equation~(\ref{dist_n}), $via$ $\s$ 
and use the corresponding $P_{\s}(\theta)$ with an isotropic hard-sphere-like distribution of the MNP inside the 
cylindrical aggregates. This textured MNP distribution will be denoted by $\rho_{hs}(\vec{r},\s)$ in the following.

\section {results}
\label {results}
We do not focus on a specific MNP assembly but instead 
in a more general way on aggregates made of standard MNP for hyperthermia applications
such as iron oxide MNP.
As in our preceding work~\cite{morjane_2025}, we use 
the example described in Ref.~\onlinecite{de-toro_2013b} to quantify the TQMC scheme. This leads to
$\la_u=6$, $T_0=97K$ and then $R_0=0.316$, $\gD{}t_{MC}=R_0^2N_p\Gamma_{acc}=3.\;10^3$ 
per period to simulate maghemite MNP of $d=8\;nm$, $K_u=3\;10^4Jm^{-3}$ at a frequency of 100\;kHz. 
However, we can extend the above set to $T_0\simeq{}300K$ in order to represent maghemite MNP 
of $c.a.~12nm$ in diameter.
The simulations presented in the following are performed with $p=8$ periods and 100 independent realizations.

First of all, we examine the effect of the DDI on both $\chi^{}_1$ and the SAR on a spherical
aggregate with a hard-sphere-like distribution of MNP and a random distribution of 
easy axes. Indeed, for the spherical-shaped aggregates in the absence of an external field, neither
a structural anisotropy nor a texturation of the easy-axis distribution is expected.
Such spherical aggregates are representative of the ones formed experimentally in the absence of an external field 
as the so-called nano-flowers,
or multicore systems, although we consider here systems including c.a. 1000 MNP.
As in the other systems considered in this work, the volume fraction of the spheres, including the coating layer, 
is fixed at $\Phi=0.45$.

\begin {figure}
       \hskip -0.01\textwidth
       \includegraphics [width = 0.38\textwidth]{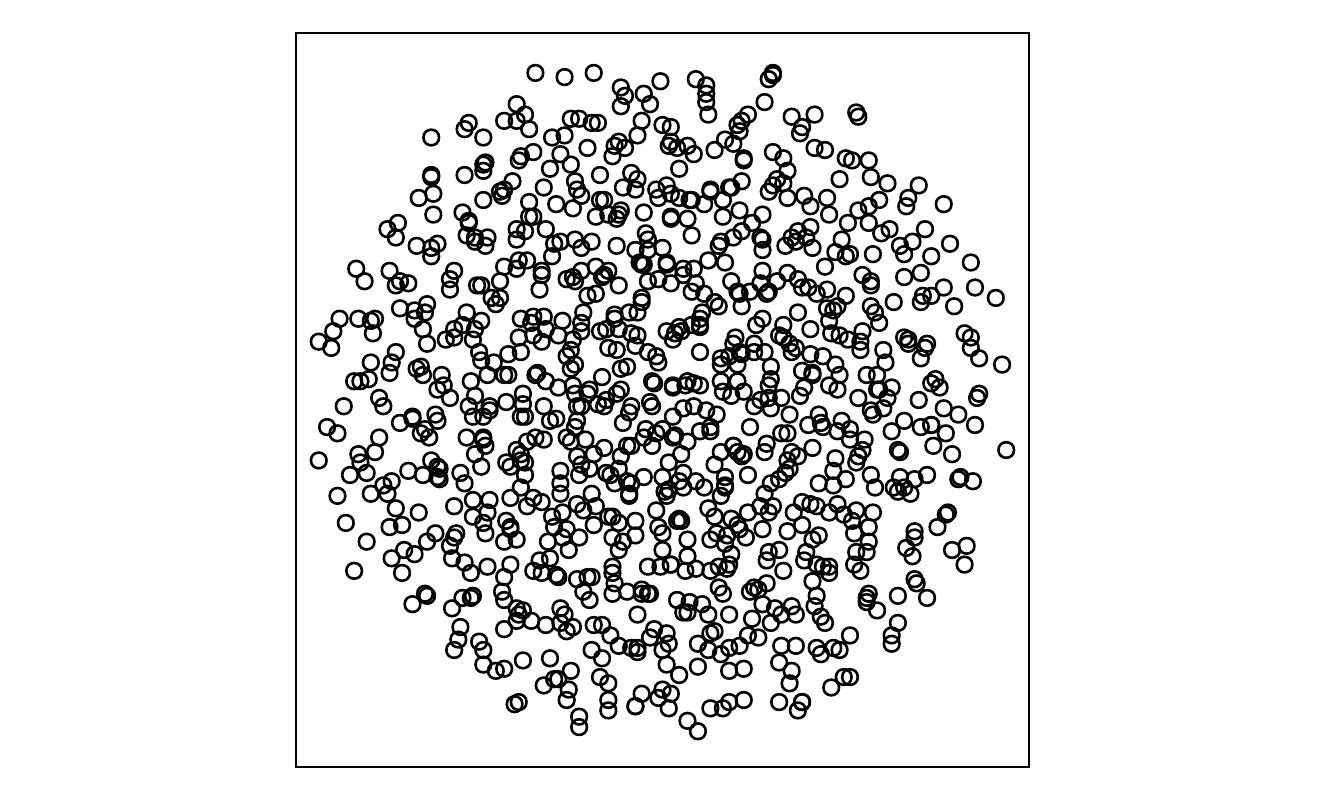}
       \hskip -0.04\textwidth
       \includegraphics [width = 0.66\textwidth]{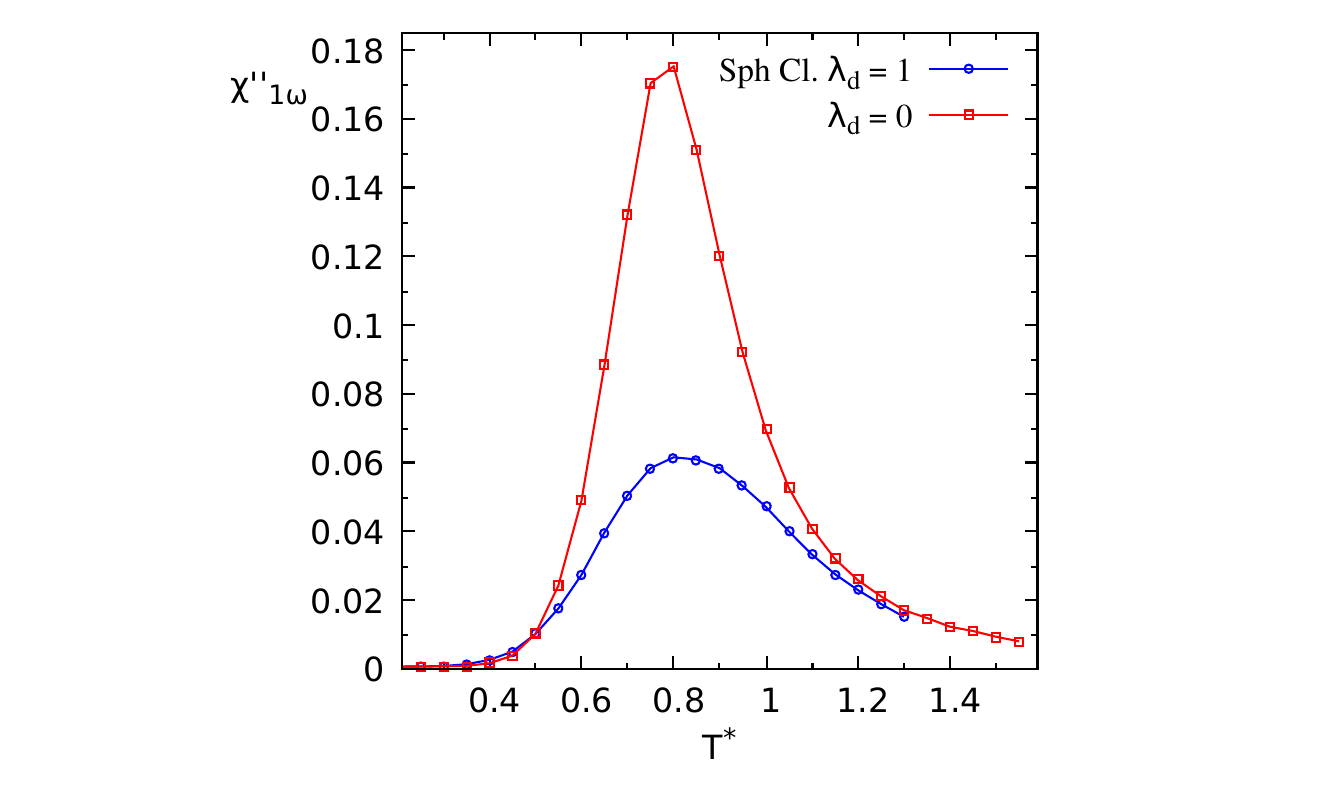}
    \caption {\label {sph_x2h05}
    The spherical cluster including 990 MNP obtained from a large cubic system
    with a hard-sphere-like distribution for 
    $\Phi=0.45$, left and the corresponding $\chi_{1\om}^{''}$ in terms of $T^*$ for $h=0.5$,
    compared to that of the isolated particles, $\la_d=0$.
    }
   \end {figure}

\begin {figure}
    \includegraphics [width = 0.66\textwidth]{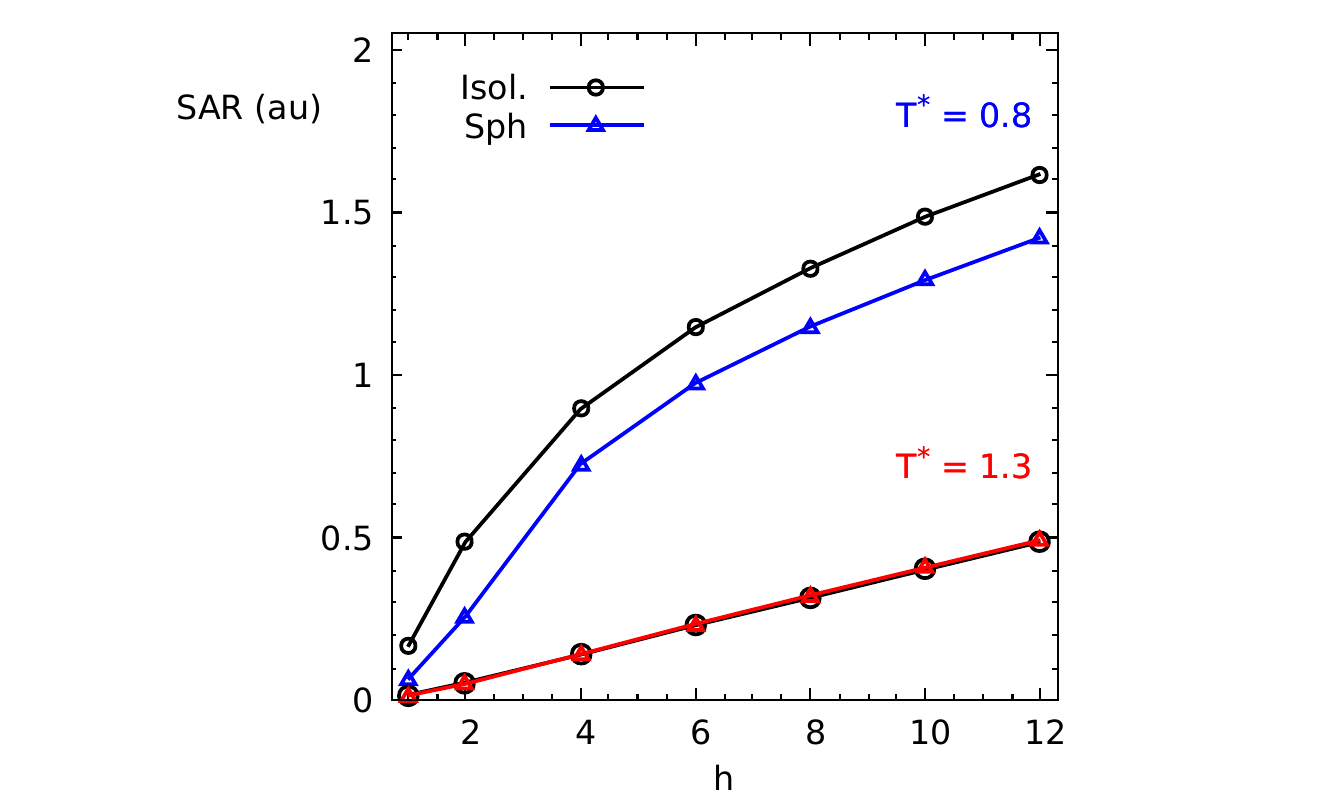}
    \caption {\label {sar_is_sph}
    SAR in reduced units ($h^2\chi_{1\om}^{''}$) in terms of $h$ at $T^*=0.8$, top and 1.30, 
    bottom for the spherical cluster of figure~(\ref{sph_x2h05}) compared to that of the 
    isolated MNP.
    }
   \end {figure}

In figure~(\ref{sph_x2h05}) we compare the $\chi^{''}_1(T^*)$ between the interacting MNP in the spherical
assembly, to that of isolated MNP, corresponding to $\la_d=0$ in terms of $T^{*}$ at a small value, $h=0.5$
of the AC field amplitude. Clearly, especially in the range of temperatures close to its maximum, the DDI induce a strong
decrease of $\chi^{''}$ and consequently of the SAR. The comparison of the SAR in terms of the external field $h$
beyond the small field range,
between the isolated MNP and the spherical cluster is shown in figure~(\ref{sar_is_sph}) at $T^*=0.8$ and 1.30. This confirms
the strong decrease of the SAR due to the DDI in the case of the isotropic spherical cluster,
except when the temperature increases beyond a threshold value where $\chi_1$ does not depend any more of $T^*$ 
(see figure ~(\ref{sph_x2h05})). 

\begin {figure}
   \hskip -0.13\textwidth
   \includegraphics [width = 0.60\textwidth]{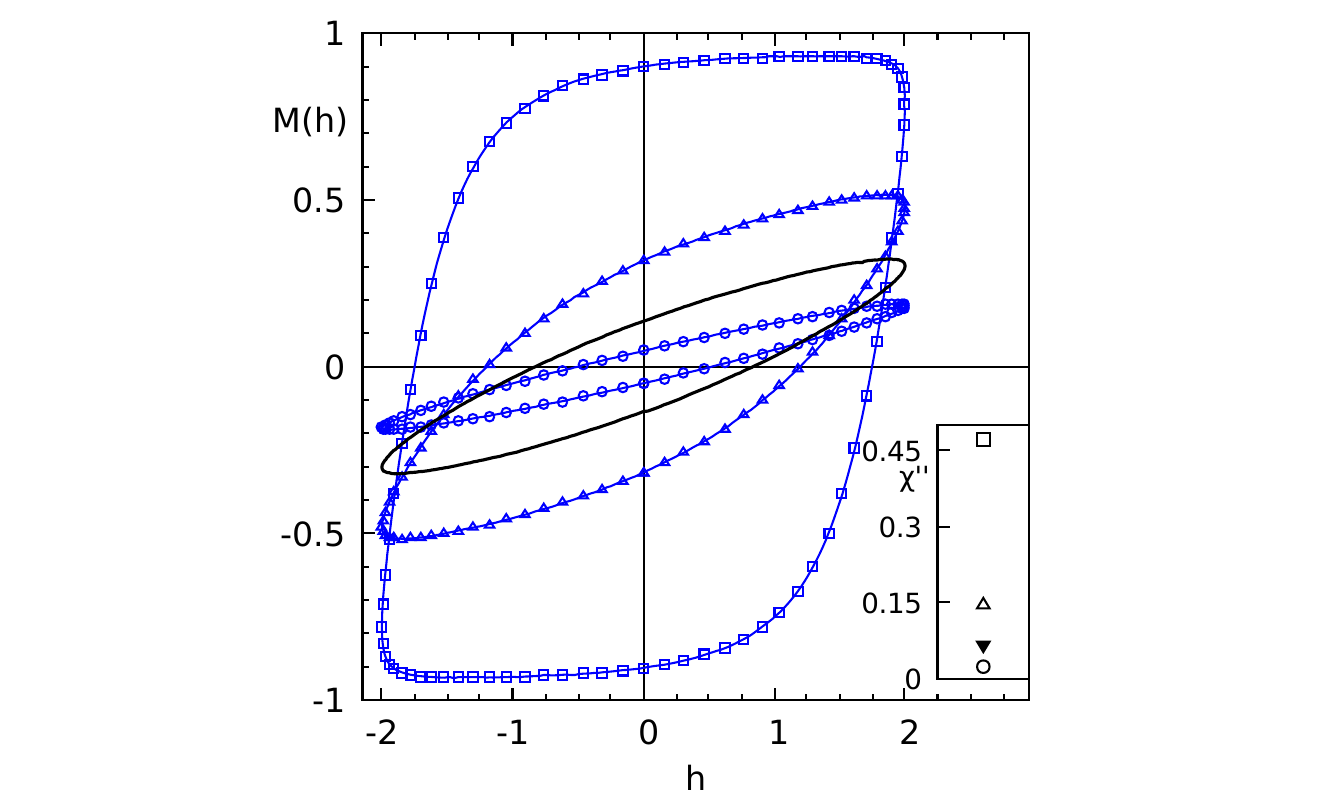}
   \hskip -0.13\textwidth
   \includegraphics [width = 0.60\textwidth]{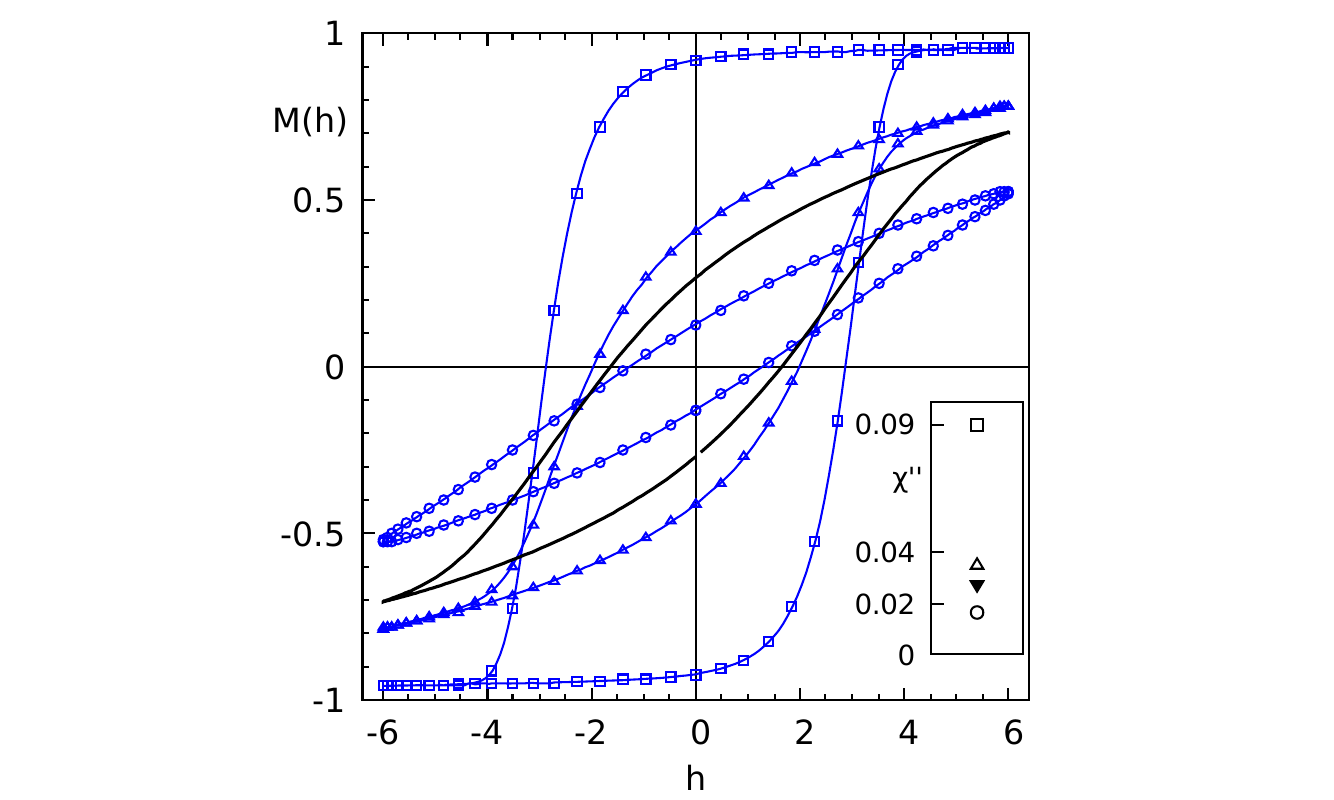}
   \caption {\label {plan_hs_sph}
   Comparison of the hysteresis curve for the slab geometry with 2D periodic boundary conditions
   to that of the spherical cluster for $h=2$, left and $h=6$, right at $T^*=0.8$ with hard sphere 
   distribution of the MNP with $\Phi=0.45$. Solid line: spherical cluster and random distribution 
   of easy axes $\{\hat{n}\}$;
   open circles: AC field along the surface normal and random distribution $\{\hat{n}\}$;
   open triangles: AC field along the surface and random distribution $\{\hat{n}\}$;
   open squares: AC field along the surface and textured distrution $\{\hat{n}\}$
   according to equation~(\ref{dist_n}) with $\s=0.2$ along the direction of the field $\hat{h}$. 
   Inset: component $\chi_{1\om}^{''}$ for the different situations; filled downward
   triangle: spherical cluster.
   }
   \end {figure}

The susceptibility and thus the SAR can be increased through the geometry of the sample together with a convenient orientation of 
the AC field. The simplest way is to use a sample with a film-like geometry,
or a slab, and to consider an exciting field parallel to the slab plane. 
Here, on figure~(\ref{plan_hs_sph}) we present a representative case with both a small $h=2$ and an intermediate value 
of the field, $h=6$ with 2D periodic boundary conditions along the plane directions $(\hat{y},\hat{z})$.
Doing this, we indeed increase the out-of-phase response, $\chi_{1}^{''}$ (see inset of figure~(\ref{plan_hs_sph})) 
but only in a rather weak amplitude, at least if the easy axes keep the random distribution. Then, by imposing a 
strongly textured distribution of the axes $\{\hat{n}_i\}$ from equation~(\ref{dist_n}) with $\s=2$, 
we can get a strong increase in the hysteresis area and accordingly in the SAR. 
However, imposing the distribution of easy axes in such a way is artificial. 

Then we use both a more singular external shape for the aggregate, namely a columnar one as a 
generalization of the one-diameter-width MNP chains and presenting, 
as explained above in section~(\ref{struct}) both a self consistent textured distribution $\{\hat{n}_i\}$ 
of the easy axes and the related anisotropic distribution of the MNP stemming from the liquid-state ferrofluid.

\begin {figure}
      \includegraphics [width = 0.66\textwidth]{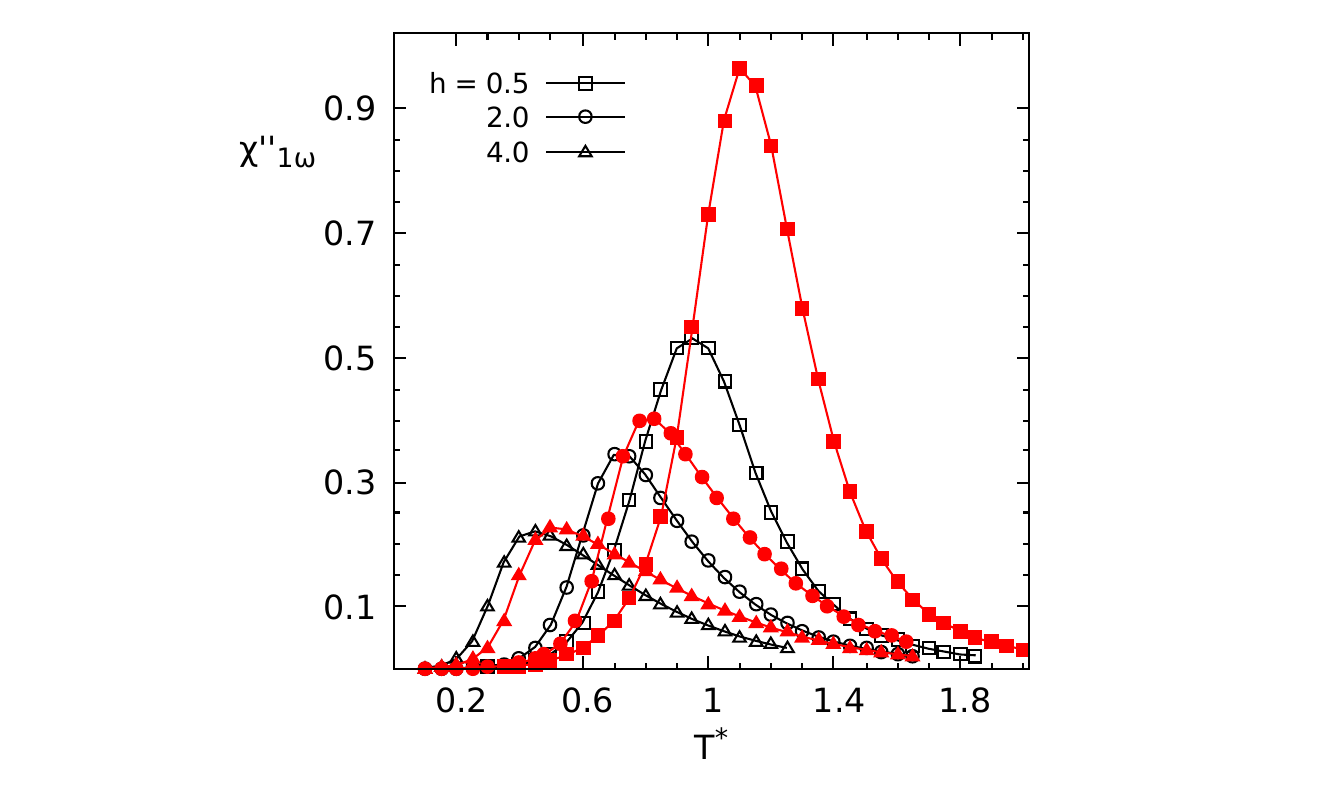}
    \caption {\label {x2_dhs_ths}
    Out of phase component of $\chi_{1\om}$ in terms of $T^*$ for the cylinder with 
    the MNP structure of :
    the DHS frozen in $h_{DC}=1$ and $\be_{fr}=0.316$ with $\{\hat{n}\}_{\infty}$ (filled symbols); 
    the hard sphere one $\rho_{hs}(\vec{r},\s)$ with $\s=0.455$ (open symbols).
    Different values of the external AC field $h$ as indicated.
    }
   \end {figure}

\begin {figure}
    \includegraphics [width = 0.66\textwidth]{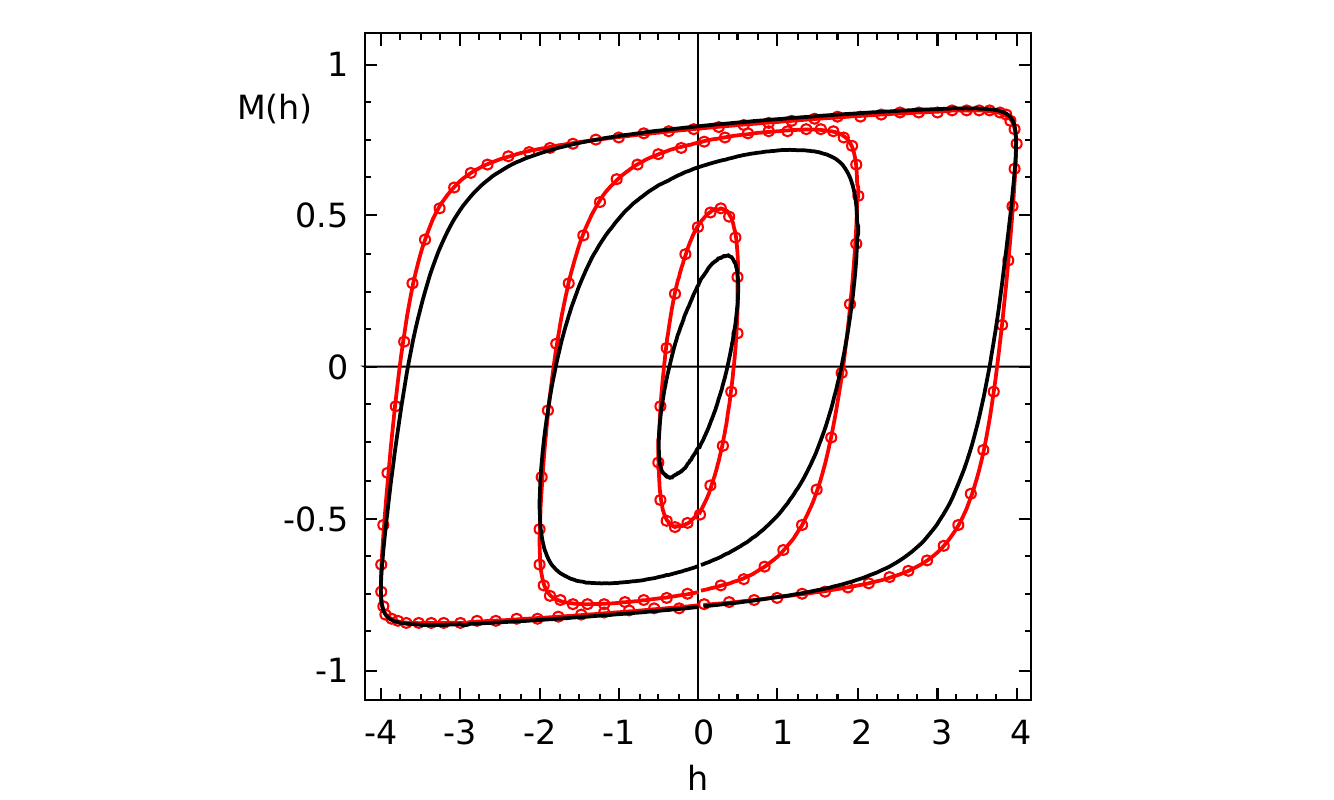}
    \caption {\label {hyst_dhs_ths}
    Hysteresis curves of the systems of figure~(\ref{x2_dhs_ths}), 
    for the DHS structure (open circles) and the HS structure
    $\rho_{hs}(\vec{r},\s=0.455)$ (solid lines)
    at the temperatures corresponding to the respective maximum of $\chi_{1}^{''}$ 
    namely $T^*=(0.53,0.83,1.10)$ for the frozen DHS 
    and $T^*=(0.45,0.70,0.95)$ for $\rho_{hs}(\vec{r},\s)$ with $h=0.5$, 2 and 4 respectively.
    }
   \end {figure}

First, we show that the structural anisotropy favors the increase of the SAR by comparing the imaginary part 
$\chi_1^{''}$ of the first harmonic susceptibility as determined using either the self-consistent frozen DHS 
structure or the hard sphere isotropic structure with the fitted textured easy axes distribution, 
$\rho_{hs}(\vec{r},\s)$.
Doing this, we compare $\chi_{1}^{''}$ between two systems presenting very similar $\{\hat{n}\}$ and differing 
only on their structure, namely the positional distribution of MNP. 
This is done on figure~(\ref{x2_dhs_ths}) in the case of the DHS frozen in an external field $h_{DC}=1$ 
at $T^{*}_{fr}=0.316$ for different values of the external field. 
Notice that in this case the difference between the self-consistent easy axes distribution $\{\hat{n}\}$ and
its $\la_u\tend\infty$ limit $\{\hat{n}\}_{\infty}$ 
is not significant, as can be seen in figure~(\ref{dist_n_be316}).
Clearly, as expected, the influence of the DDI-induced anisotropy is all the more important when the field is small. 
The DDI-induced structural anisotropy has two effects. 
Firstly, the temperature $T_m^{*''}$ of the maximum of $\chi_1^{''}(T^*)$ is increased
and secondly the value of $\chi_1^{''}(T^*)$ is increased at temperatures $T^*$ greater than the value of 
$T_m^{*''}$ as determined with the isotropic structure $\rho_{hs}(\vec{r},\s)$. 
Thus we see that both the DDI-induced structural anisotropy and the texturation of the easy axes distribution 
play a noticeable role in the increase of $\chi_1^{''}$. This is in agreement with the behavior of the bulk 
frozen DHS~\cite{malherbe_2023}.
The hysteresis curves at the respective temperatures $T_m^{*''}$ are shown in figure~(\ref{hyst_dhs_ths}).
We clearly see that, except at high values of the external field, the shift of the value of $T_m^{*''}$
is not sufficient to make the hysteresis of the system coincide with that of an isotropic structure or with that of a 
true DHS-like structure even with the same texturation of the easy axes.  
Therefore, we show that the intrinsic DDI-induced anisotropy of the structure in position is an important 
element in the correct determination of the columnar aggregates susceptibility.

\begin {figure}
    \hskip -0.13\textwidth
    \includegraphics [width = 0.60\textwidth]{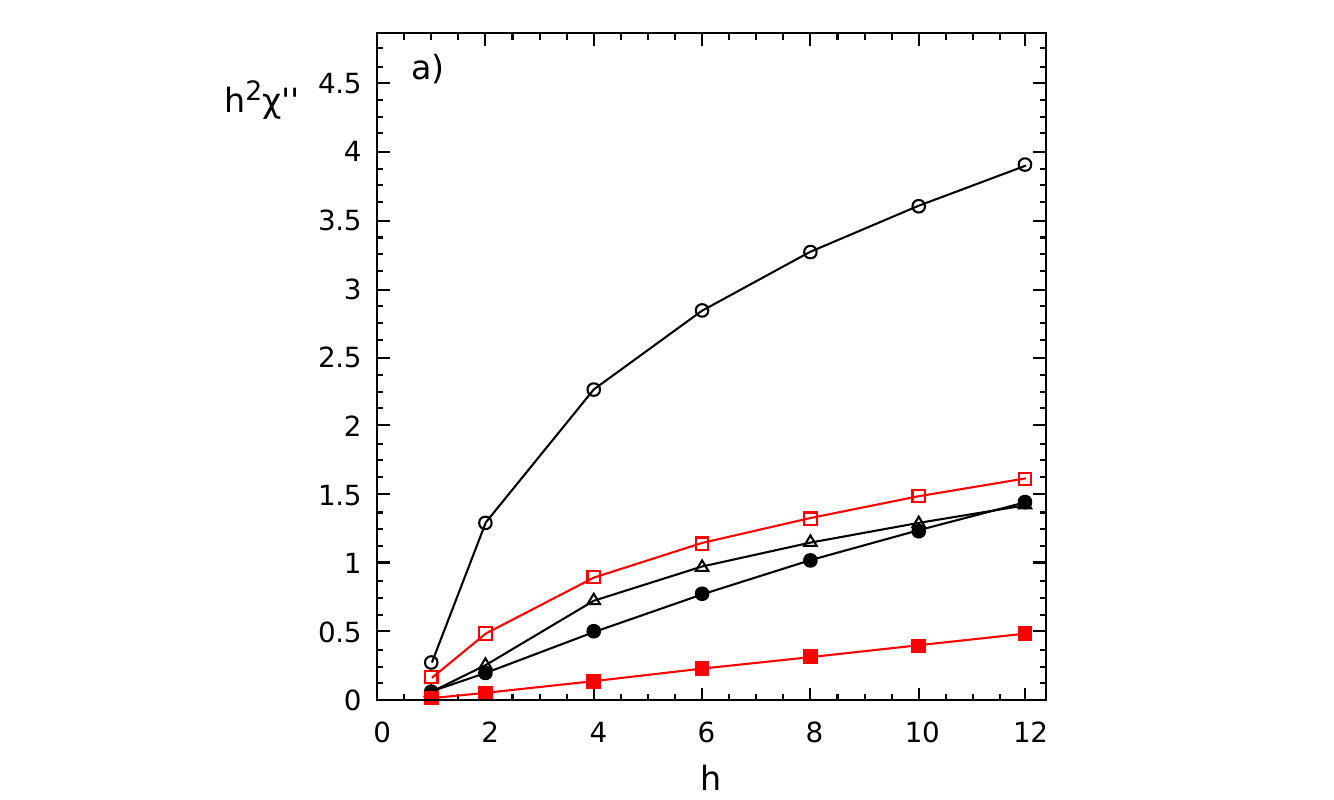}
    \hskip -0.13\textwidth
    \includegraphics [width = 0.60\textwidth]{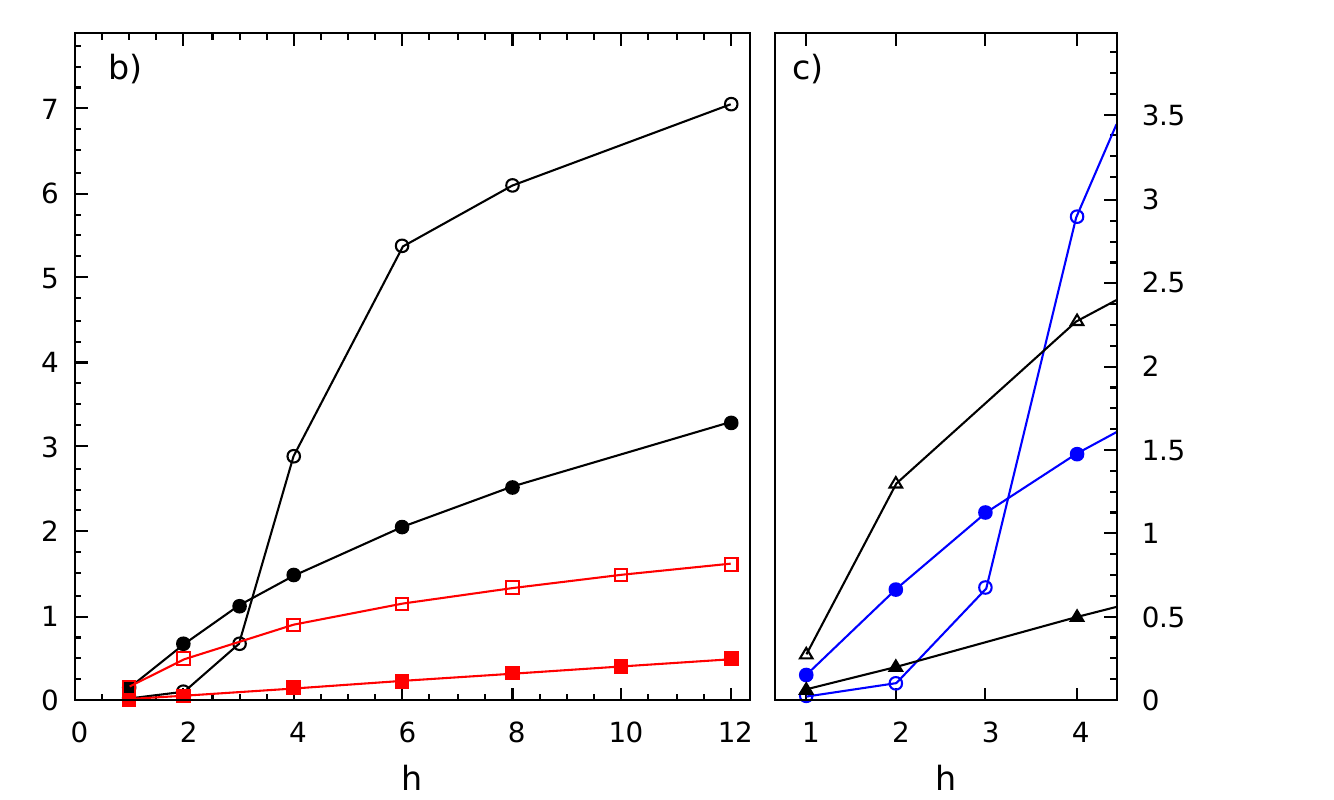}
   \caption {\label {sar_ed4}
   Reduced SAR ($h^2\chi_{1\om}^{''}$) for the cylindrical aggregate in terms of the field with the DHS structure
   frozen in presence of the DC field $h_{DC}=|h_{AC}|=h$ for a): $\ep_d=1$ and b): $\ep_d=4$, 
   with $T^*=0.8$ (open circles) and 1.30 (filled circles), and comparison with the
   isolated ensemble, $\ep_d=0$, (open squares) and (filled squares) respectively. 
   In the case $\ep_d=1$ we also compare to the result of the spherical aggregate with 
   isotropic structure and random distribution of easy axes (open triangles). The latter 
   is only given at $T^*=0.8$ since at $T^*=1.30$ the result for $\chi^{''}_{1\om}$ nearly 
   coincides with its $\ep_d=0$ counterpart.
   c) Comparison of the reduced SAR for $\ep_d=1$, (triangles) and $\ep_d=4$, (circles) at
   $T^*=0.8$ (open) and 1.30 (filled) symbols respectively.
   }
  \end {figure}

Finally, we simulate the case where the structure of the system in the columnar aggregate results from
the AC field itself. As we mentioned in section~(\ref{struct}), we model the structure resulting from the 
AC field $|\vec{h}_{AC}|=h$ of the hyperthermia process by that obtained in a DC field along the cylinder 
with $h_{DC}=h$. Then this structure is frozen, and we perform the dynamical simulations on the cylindrical
aggregate characterized by this frozen structure.
In figure~(\ref{sar_ed4}) we present the SAR in terms of $h$ for such a structure of the cylindrical aggregate
at the two characteristic values of the temperature, $T^{*}=0.8$ and 1.30. The latter corresponds to the situation
where the component $\chi_{1}^{''}$ becomes nearly independent of the DDI, while conversely the former 
corresponds to a peak of the DDI effect at low fields (see figure~(\ref{sph_x2h05})).
As expected from the result in the planar geometry with strongly textured easy axes distribution,
we clearly see that, in contrast to what we get on an isotropic spherical cluster, the DDI strongly
enhance the SAR, at least at $T^{*}=0.8$. 
The other result here is the behavior with respect to the value of $\ep_d$ when this latter
increases beyond $\ep_d=1$. Indeed, in figure~(\ref{sar_ed4}c) we clearly observe an evidence of a DDI-induced frozen 
behavior for low values of $T^{*}$ since the SAR is an increasing fonction of $\ep_d$ only beyond $h$ 
values allowing to get rid of this freezing behavior. In the particular case displayed in 
figure~(\ref{sar_ed4}c) this happens for $h>3.5$, corresponding to c.a. $H_{K}/3.5$, at $T^*=0.8$.
The precise investigation of this freezing behavior, related to the spin-glass behavior, is beyond the
scope of the present work.

\section {Concluding remarks}
\label {conclusion}

In this work, a Time Quantified Monte Carlo framework is used in order
to study the DDI influence on the hyperthermia efficiency through the value of the
SAR on MNP aggregates differing by their external shape and local structure. 
According to Ref.~\onlinecite{barrera_2023}, we point out that the SAR, defined on the complete
period of the AMF and when the latter is of sine form, is determined by the
imaginary component of the first harmonic response, $\chi_{1\om}$, not to be
confused with the linear response. 
We first showed that for an isotropic spherical-shaped aggregate, the DDI decrease the SAR.
In the film-like geometry, as expected, the SAR increased with respect to the case of isolated MNP
with an AMF along the surface plane
all the more that the easy axes are textured in the direction of the AMF.
Then in columnar aggregates, which are represented as an assembly of MNP constrained to
a cylindrical simulation box, we obtain a strong enhancement of the SAR when the AMF
direction is along the cylinder axis.
We emphasize that the local structure of the MNP in the aggregate must be obtained from
the equilibrium distribution of the corresponding ferrofluid before freezing in an immobilized
MNP distribution.
This equilibrium step was performed with a parallel-tempering MC scheme.
The important results of our work are, first, that the DDI-induced anisotropy of the local
structure plays an important role in the SAR increase, and this is in agreement with a 
previous result of the DDI-induced collective behavior of the bulk dipolar hard sphere 
fluid~\cite{malherbe_2023}, and second that the easy axes distribution must be determined
in a self-consistent way from the MNP structure.

Finally we note that our results concerning the increase of the SAR from an isotropic
spherical and a cylindrical aggregate as seen in figure~(\ref{sar_ed4}a) is in qaulitative agreement
with the result of Ref.~\onlinecite{gandia_2019} where a similar comparison is done
experimentally between the SAR of magnetotactic bacteria to that of the corresponding
magnetosomes arranged in a spherical-like aggregate.

\section {Acknowledgements}
 This work was granted access to the HCP resources of CINES under allocations
 2025-A0180906180 and 2026-A0200906180 made by GENCI, CINES, France.
 We acknowledge the support of the French Agence Nationale de la Recherche (ANR) through the project
 ANR NanoHype under grant number ANR-21-CE09-0043.
 Important discussions with I. Lisiecki, A.-T. Ngo and A. Shahmanesh from MONARIS, CNRS and Sorbonne Université
 and with H. Kachkachi from PROMES, CNRS and Université de Perpignan are greatly acknowledged. 
    \FloatBarrier
  %
  %
   %
   %
   %
   %
   %
   %
   %
   %
 %
    \FloatBarrier

\begin{thebibliography}{10}

\bibitem{pankhurst_2009}
Q.A. Pankhurst, N.K.T. Thanh, S.~K. Jones, and J.~Dobson.
\newblock {\em J. Phys. D}, 42:224001, 2009.

\bibitem{gavilan_2021b}
H.~Gavil\'{a}n, S.~K. Avugadda, T.~Fern\'{a}ndez-Cabada, a~N.~Soni, M.~Cassani,
  B.~T. Mai, R.~Chantrell, and T.~Pellegrino.
\newblock {\em Chem. Soc. Rev.}, 50:11614, 2021.

\bibitem{baldea_2025}
I.~Baldea, C.~Iacovit, R.A. Gurgu, A.S. Vizitiu, V.~Râzniceanu, and D.~R.
  Mitrea.
\newblock {\em Nanomaterials}, 15:1519, 2025.

\bibitem{bordet_2016}
A.~Bordet, L.-M. Lacroix, P.-F. Fazzini, J.~Carrey, K.~Soulantica, , and
  B.~Chaudret.
\newblock {\em Angew. Chem. Int. Ed.}, 55:15894, 2016.

\bibitem{bordet_2025}
A.~Bordet, W.~Leitner, , and B.~Chaudret.
\newblock {\em Angew. Chem. Int. Ed.}, 64:e202424151, 2025.

\bibitem{hergt_2006a}
R.~Hergt, S.~Dutz, R.~Müller, and M.~Zeisberger.
\newblock {\em J. Phys.: Condens. Matter}, 18:S2919, 2006.

\bibitem{keblinski_2006}
P.~Keblinski, D.G. Cahill, A.~Bodapati, C.R. Sullivan, and T.A. Taton.
\newblock {\em J. Appl. Phys.}, 100:054305, 2006.

\bibitem{eggeman_2007}
A.~S. Eggeman, S.~A. Majetich, D.~Farrell, and Q.~A. Pankhurst.
\newblock {\em IEEE Trans. Magn.}, 43:2451, 2007.

\bibitem{dejardin_2022}
J.-L. Déjardin and H.~Kachkachi.
\newblock {\em J. Magn. Magn. Mat.}, 556:169354, 2022.

\bibitem{kachkachi_2026}
H.~Kachkachi.
\newblock {\em J. Appl. Phys.}, 2026.
\newblock In press; arxiv:2606.04724v1.

\bibitem{jonsson_2001}
P.E. Jonsson and J.L. Garc\'{i}a-Palacios.
\newblock {\em Phys. Rev. B}, 64:174416, 2001.

\bibitem{azeggagh_2007}
M.~Azeggagh and H.~Kachkachi.
\newblock {\em Phys. Rev. B}, 75:174410, 2007.

\bibitem{haase_2012}
C.~Haase and U.~Nowak.
\newblock {\em Phys. Rev. B}, 85:045435, 2012.

\bibitem{serantes_2010}
D.~Serantes, D.~Baldomir, C.~Martinez-Boubeta, K.~Simeonidis, M.~Angelakeris,
  E.~Natividad, M.~Castro, A.~Mediano, D.-X. Chen, A.~Sanchez, LI. Balcells,
  and B.~Martínez.
\newblock {\em J. Appl. Phys.}, 108:073918, 2010.

\bibitem{mehdaoui_2013}
B.~Mehdaoui, R.P. Tanand~A. Meffre, J.~Carrey, S.~Lachaize, B.~Chaudret, , and
  M.~Respaud.
\newblock {\em Phys. Rev. B.}, 87:174419, 2013.

\bibitem{gudoschnikov_2012a}
S.A. Gudoshnikov, B.Ya. Liubimov, A.V. Popova, and N.A. Usov.
\newblock {\em J. Magn. Magn. Mat.}, 324:3690, 2012.

\bibitem{gudoschnikov_2012b}
S.~A. Gudoshnikov, B.~Ya. Liubimov, , and N.~A. Usov.
\newblock {\em Aip Adv.}, 2:012143, 2012.

\bibitem{sanz_2020}
B.~Sanz, R.~Cabreira-Gomes, T.~E. Torres, D.~P. Valdés, Jr. E.~Lima, E.~De
  Biasi, R.D. Zysler, M.R. Ibarra, , and G.F. Goya.
\newblock {\em Acs Appl. Nano Mat.}, 3:8719, 2020.

\bibitem{morales_2021}
I.~Morales, R.~Costo, N.~Mille, J.~Carrey, A.~Hernando, and P.~de~la Presa.
\newblock {\em Nanoscale Adv.}, 3:5801, 2021.

\bibitem{mille_2021}
S.~Faure J. M. Asensio a) B.~Chaudret N.~Mille, D. De~Masi and J.~Carrey.
\newblock {\em Appl. Phys. Lett.}, 119:022407, 2021.

\bibitem{arica_2026}
T.A. Arica, E.~Tiryaki, E.~Sadeghi, A.R. Mannir, S.~Balci, and T.~Pellegrino.
\newblock {\em nanoscale}, 18:4773, 2026.

\bibitem{gandia_2019}
D.~Gandia, L.~Gandarias, I.~Rodrigo, J.~Robles-García, R.~Das, E.~Garaio,
  J.~Á. García, M.-H. Phan, H.~Srikanth, I.~Orue, J.~Alonso, A.~Muela, and
  M.~L. Fdez-Gubieda.
\newblock {\em Small}, 15:1902626, 2019.

\bibitem{martinez-boubeta_2013}
C.~Martinez-Boubeta, K.~Simeonidis, A.~Makridis, M.~Angelakeris, O.~Iglesias,
  P.~Guardia, A.~Cabot, L.~Yedra, S.~Estradé, F.~Peiró, Z.~Saghi, P.~A.
  Midgley, I.~Conde-Leborán, D.~Serantes, and D.~Baldomir.
\newblock {\em Sci. Rep.}

\bibitem{anand_2019}
M.~Anand, V.~Banerjee, and J.~Carrey.
\newblock {\em Phys. Rev. B}, 99:024402, 2019.

\bibitem{anand_2020}
M.~Anand.
\newblock {\em J. Appl. Phys.}, 128:023903, 2020.

\bibitem{anand_2022}
M.~Anand.
\newblock {\em Pramana J. Phys.}, 96:132, 2022.

\bibitem{saville_2014}
S.~L. Saville, B.~Qi, J.~Baker, R.~Stone, R.~E. Camley, K.~L. Livesey, L.~Ye,
  T.~M. Crawford, and O.~T. Mefford.
\newblock {\em J. Coll. Interf. Sci.}, 424:141, 2014.

\bibitem{schoenen_2025}
M.~Schoenen, T.~Schmitz-Rode, and I.~Slabu.
\newblock {\em Computer Methods and Programs in Biomedicine}, 267:108775, 2025.

\bibitem{morjane_2025}
A.~Morjane, J.-G. Malherbe, Juan~J. Alonso, F.~Vernay, and V.~Russier.
\newblock {\em Journal of Magnetism and Magnetic Materials}, 630:173443, 2025.

\bibitem{malherbe_2023}
J.-G. Malherbe, J.J. Alonso, and V.~Russier.
\newblock {\em J. Phys. Condens. Matt.}, 35:05802, 2023.

\bibitem{allen_1987}
M.~P. Allen and D.~J. Tildesley.
\newblock {\em Computer Simulation of Liquids}.
\newblock Oxford Science Publications, 1987.

\bibitem{weis_1993}
J.-J. Weis and D.~Levesque.
\newblock {\em Phys. Rev. E}, 48:3728, 1993.

\bibitem{earl_2005}
D.-J. Earl and M.-W. Deema.
\newblock {\em Phys. Chem. Chem. Phys.}, 7:3910, 2005.

\bibitem{nowak_2000}
U.~Nowak, R.~W. Chantrell, and E.~C. Kennedy.
\newblock {\em Phys. Rev. Lett.}, 84:163, 2000.

\bibitem{cheng_2006}
X.~Z. Cheng, M.~B.~A. Jalil, and H.~K. Lee.
\newblock {\em Phys. Rev. B}, 73:224438, 2006.

\bibitem{andersson_1997}
J.-O. Andersson, C.~Djurberg, T.~Jonsson, P.~Svedlindh, and P.~Nordblad.
\newblock {\em Phys. Rev. B}, 56:13983, 1997.

\bibitem{barrera_2023}
G.~Barrera, P.~Allia, and P.~Tiberto.
\newblock {\em J. Phys. Materials}, 6:035007, 2023.

\bibitem{rosensweig_2002}
R.E. Rosensweig.
\newblock {\em J. Magn. Magn. Mat.}, page 370, 2002.

\bibitem{de-toro_2013b}
J.~A. De-Toro, P.~S. Normile, S.~S. Lee, D.~Salazar, J.~L. Cheong, P.~Muniz,
  J.~M. Riveiro, M.~Hillenkamp, F.~Tournus, A.~Tamion, and P.~Nordblad.
\newblock {\em J. Phys. Chem. C}, 117:10213, 2013.

\bibitem{garcia-acevedo_2023}
P.~García-Acevedo, M.~González-Gómez, A.~Arnosa-Prieto, L.~de~Castro-Alves,
  Y.~Piñeiro, and J.~Rivas.
\newblock {\em Adv. Sci.}, 10:2203397, 2023.

\bibitem{levesque_2014}
D.~Levesque and J.-J. Weis.
\newblock {\em J. Chem. Phys.}, 140:094507, 2014.

\bibitem{chantrell_2025}
R.~Chantrell.
\newblock 2025.
\newblock Private communication.

\end{thebibliography}

\end   {document}